\documentclass[aps,prb,twocolumn,superscriptaddress,nopacs,amsmath,amssymb]{revtex4}

\usepackage{graphicx}
\usepackage{epstopdf}

\usepackage[version=3]{mhchem} 
\usepackage{color}

\begin{document}

\title{Superconductivity in Tl$_{0.6}$Bi$_{2}$Te$_{3}$ Derived from a Topological 
Insulator\footnote{This document is the unedited authors' version of a submitted work that was subsequently accepted for publication in Chemistry of Materials (copyright American Chemical Society) after peer review. To access the final edited and published work see the journal's website.}}


\author{Zhiwei~Wang}

\affiliation{Institute of Physics II, University of Cologne, D-50937 Cologne, Germany}
\affiliation{Institute of Scientific and Industrial Research, Osaka University, Osaka 567-0047, Japan}


\author{A.~A.~Taskin}

\affiliation{Institute of Physics II, University of Cologne, D-50937 Cologne, Germany}
\affiliation{Institute of Scientific and Industrial Research, Osaka University, Osaka 567-0047, Japan}

\author{Tobias~Fr\"olich}

\author{Markus~Braden}
\affiliation{Institute of Physics II, University of Cologne, D-50937 Cologne, Germany}

\author{Yoichi~Ando}
\email{ando@ph2.uni-koeln.de}

\affiliation{Institute of Physics II, University of Cologne, D-50937 Cologne, Germany}
\affiliation{Institute of Scientific and Industrial Research, Osaka University, Osaka 567-0047, Japan}

\date{\today}


\begin{abstract}

Bulk superconductivity has been discovered in Tl$_{0.6}$Bi$_2$Te$_3$,
which is derived from the topological insulator Bi$_2$Te$_3$. 
The superconducting volume fraction of up to 95\%
(determined from specific heat) with ${T_c}$ of 2.28 K was observed.
The carriers are $p$-type with the density 
of $\sim$1.8 $\times$ 10$^{20}$ cm$^{-3}$. 
Resistive transitions under magnetic fields
point to an unconventional temperature dependence of the upper critical field $B_{c2}$. 
The crystal structure appears to be unchanged from Bi$_2$Te$_3$ with 
a shorter $c$ lattice parameter, which, together with the Rietveld analysis, 
suggests that Tl ions are incorporated but not intercalated.
This material is an interesting candidate of a topological superconductor
which may be realized by the strong spin-orbit coupling inherent to
topological insulators.

\end{abstract}

\maketitle

\section{Introduction}

The advent of topological insulators have created an exciting
interdisciplinary research field which is vitalized by discoveries of new
materials to realize new concepts \cite{Zhang_Science_07, Fu_PRL_07,
Moore_PRB_07, Hsieh_Nature_08, Roy_PRB_09, Alexey_PRB_09, Sato_PRL_10,
Hasan_RMP_10, Qi_RMP_11, Ando_JPSJ_13} and hence is greatly helped by
contributions from chemistry.\cite{CavaReview} The topological
insulators are characterized by a gapped bulk state and gapless surface
or edge states whose gapless nature is protected by time-reversal
symmetry. Soon after the discovery of topological insulators, it was
recognized that a similar topological state is conceivable for
superconductors which also have a gapped bulk state.\cite{Schnyder}
Already various schemes for realizing such a topological
superconductor (TSC) have been discussed,\cite{Fu_PRL_08, Qi_PRL_09,
Sato_PRB_10, Linder_PRL_10} inspired by the interest in exotic quasiparticles 
called Majorana fermions which may show up in TSCs. \cite{Wilczek_09}
In particular, it has
been proposed \cite{Fu-Berg} that superconductors derived from
topological insulators are promising candidates of TSCs due to the
strong spin-orbit coupling which would lead to unconventional electron pairing.
For superconductors of this category, a
limited number of materials, such as
Cu$_x$Bi$_2$Se$_3$,\cite{Hor_PRL_10, Markus_PRL_11} Bi$_2$Te$_3$ under
high pressure,\cite{Zhang_HP-Bi2Te3_PNAS}
In$_x$Sn$_{1-x}$Te,\cite{Sasaki_PRL_12}
Cu$_x$(PbSe)$_5$(Bi$_2$Se$_3$)$_6$,\cite{Sasaki_PRB_14} 
Sr$_x$Bi$_2$Se$_3$,\cite{SrxBi2Se3_15} and Tl$_5$Te$_3$ \cite{Tl5Te3} 
have been discovered and studied.

Among such candidate TSCs, Cu$_x$Bi$_2$Se$_3$ was the first to show
intriguing signatures of Majorana fermions on the surface.\cite{Sasaki_PRL_11}
The superconductivity in this material occurs as a result of Cu
intercalation into the van der Waals gap of the parent Bi$_2$Se$_3$
compound. Although superconducting Cu$_x$Bi$_2$Se$_3$ can be grown by a melting
method,\cite{Hor_PRL_10} the superconducting volume fraction (VF) is
typically very low (up to $\sim$20\%) in melt-grown samples. It was
shown that an electrochemical synthesis technique\cite{Markus_PRB_11}
yields samples with much higher superconducting VF (up to $\sim$70\%)
near $x$ = 0.3.\cite{Markus_PRL_11} However, chemical differences 
between superconducting and nonsuperconducting samples of 
Cu$_x$Bi$_2$Se$_3$ are not understood.
The superconductor phase is apparently unstable and it is easily
lost by heat or mechanical strain, which makes it difficult to elucidate its 
exact crystal structure.

Very recently, it was found that bulk superconductivity can also be
achieved in Bi$_2$Se$_3$ by intercalation of Sr; in the resulting
Sr$_x$Bi$_2$Se$_3$, the maximum transition temperature $T_c$ of 2.9 K and the superconducting
VF of up to 90\% have been reported. \cite{SrxBi2Se3_15,
Shruti_Arxiv_15} Also, it has been reported that all the binary
topological-insulator materials having the tetradymite structure,
Bi$_2$Se$_3$, Bi$_2$Te$_3$, and Sb$_2$Te$_3$, become superconductors
under high pressure,\cite{Zhang_HP-Bi2Te3_PNAS, HP-Bi2Se3_PRL,
HP-Sb2Te3_SR} although it is still to be elucidated how the
crystallographic and electronic structures are altered before these
systems show superconductivity under pressure. Another interesting
candidate of TSC is
Sn$_{1-x}$In$_x$Te.\cite{Sasaki_PRL_12} This is derived from the
topological crystalline insulator \cite{Ando_ARCMP} SnTe by doping In
to the Sn site, after which the topological surface states are still
preserved. \cite{Sato_PRL_13} However, the topological superconducting state appears to be
limited to a narrow range of $x$ and the condition for its realization is 
not clear at the moment. \cite{Mario_PRB_13}

To foster the research of TSCs, further discoveries of candidate
materials are desirable. In this regard, making Bi$_2$Te$_3$
superconducting in ambient pressure by doping would be
very useful, because it allows for direct comparison to Cu$_x$Bi$_2$Se$_3$
or Sr$_x$Bi$_2$Se$_3$.
Like Bi$_2$Se$_3$, pristine Bi$_2$Te$_3$ consists of covalently bonded
quintuple layers (QLs) having the stacking sequence of Te-Bi-Te-Bi-Te,
and those QLs are held together by van der Waals force,
\cite{Bi2Te3_structure_1960} which is weak enough to allow for easy exfoliation.
In contrast to Bi$_2$Se$_3$ in which superconductivity is known to show
up upon intercalation of Cu or Sr, no robust superconductivity has been
reported for intercalated Bi$_2$Te$_3$, besides a preliminary report
\cite{Hor_PdBi2Te3} of a trace superconductivity in Pd$_x$Bi$_2$Te$_3$
which has not been confirmed by other groups. In this paper, we report that
doping a large amount of Tl to Bi$_2$Te$_3$ results in a 
superconductor with a transition temperature of 2.28 K. A large
superconducting VF of up to 95\% determined from specific-heat measurements
gives evidence for the bulk nature of the superconductivity in
Tl$_{0.6}$Bi$_2$Te$_3$. This discovery could provide a new 
platform for addressing topological superconductivity.

\section{Experimental Methods}

Single crystalline samples with the nominal composition of
Tl$_{x}$Bi$_2$Te$_3$ with various $x$ values were synthesized from
high-purity elemental shots of Tl (99.99\%), Bi (99.9999\%) and Te
(99.9999\%). We focus on samples with $x$ = 0.6 in this paper, and
results on other $x$ values are presented in the Supporting Information. 
Before the synthesis, we performed
surface cleaning procedures to remove the oxide layers formed in air on
the raw shots of starting materials, as described in our previous paper.
\cite{Zhiwei_Cr-TlSbTe2} The raw materials were then mixed with the
total weight of 4.0 g and sealed in an evacuated quartz tube. The sealed
quartz tubes were heated up to 1123 K and kept for 48 h with
intermittent shaking to ensure homogeneity of the melt. The tubes
were subsequently cooled down to 823 K at a rate of 5 K/h and, then, quenched into
ice-water. We also prepared a similar sample without quenching and found
that quenching is essential for obtaining superconducting samples. Large
shiny single crystals with the lateral dimension of up to a few
centimeters can be obtained by cleaving along the $ab$ plane. The
reference Bi$_2$Te$_3$ crystal was grown with the same method involving
quenching. In addition, we also synthesized Tl$_{x}$Bi$_{2-x}$Te$_3$ with exactly
the same method for comparison.

The crystal structure was analyzed with X-ray diffraction (XRD)
using $\theta$--2$\theta$ scan performed on Rigaku Ultima-IV 
X-ray apparatus. The Rietveld analyses of powder XRD data were 
performed by using FullProf software package. The actual composition was
analyzed by using inductively coupled plasma atomic-emission
spectroscopy (ICP-AES) as well as energy-dispersive X-ray spectroscopy (EDX). 
DC magnetic susceptibility was measured in a
SQUID magnetometer (Quantum Design MPMS). The in-plane transport
properties were measured by a standard six-probe method, recording the
longitudinal resistivity $\rho_{xx}$ and the Hall resistivity
$\rho_{yx}$ simultaneously. The single crystal samples for transport
measurements were cut into a rectangular shape with a typical size of $2
\times 0.5 \times$ 0.2 mm$^3$, and electrical contacts were made by
using room-temperature-cured silver paste. The specific heat $c_{p}$ was
measured by a relaxation-time method using the Physical Properties
Measurement System from Quantum Design equipped with a $^3$He probe; the
addenda signal was measured before mounting the sample and was duly
subtracted from the measured signal. The $c_{p}$ measurements were done
in 0 T as well as in various magnetic fields up to 2 T applied along the
$c$ axis.

\section{Results and discussions}


\begin{figure}[t]
\includegraphics[width=8.5cm,clip]{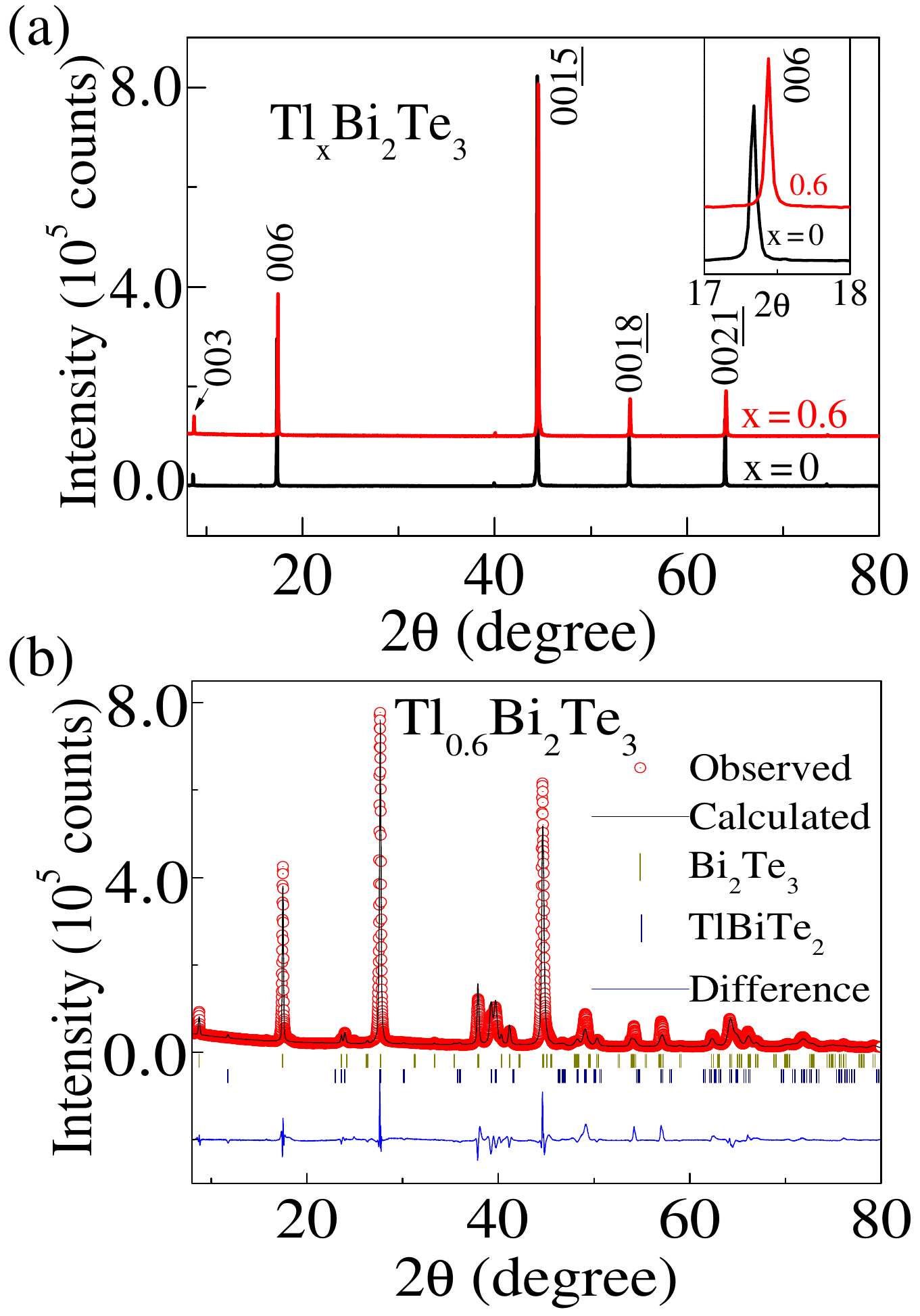}
\caption{(a) XRD patterns of Tl$_{0.6}$Bi$_2$Te$_3$ and Bi$_2$Te$_3$,
showing (00$l$) reflections from cleaved single
crystals; inset shows an enlarged view of the (006) peak, which presents
a clear shift to higher angle after Tl doping. 
(b) Powder XRD data for Tl$_{0.6}$Bi$_2$Te$_3$ 
taken on powders prepared by crushing cleaved single crystals, together with 
the result of a Rietveld refinement to consider a coexistence of Bi$_2$Te$_3$ 
and TlBiTe$_2$. 
Red symbols denote the observed intenstities; black and blue lines give the caculated 
and difference intensities, respectively. The upper and lower lines of vertical bars 
indicate the positions of the Bragg reflections of the main and impurity phases, respectively.
} 
\label{Crystal_plot}
\end{figure}

We found that quenched single crystals of Tl$_{0.6}$Bi$_2$Te$_3$ are invariably
superconducting at low temperature. This composition suggests that Tl atoms are
intercalated in the van der Waals gap of Bi$_2$Te$_3$; however, as we show in 
the following, the crystal structure analysis suggests that intercalation is {\it not} 
taking place.
Figure 1(a) shows the XRD pattern of Tl$_{0.6}$Bi$_2$Te$_3$ 
measured on cleaved single
crystals, along with similar data for pristine Bi$_2$Te$_3$. 
The sharp reflections indicate good crystalline quality of our
single crystals. Only (00$l$) reflections can be observed with this method, and the peaks
are easily indexed by considering the rhombohedral structure of
Bi$_2$Te$_3$. Hence, after the doping of Tl into
Bi$_2$Te$_3$, the crystal structure remains essentially the same as that of the
parent compound. However, in contrast to the cases of Cu- or
Sr-doped Bi$_2$Se$_3$, in which those dopants are intercalated into the
van der Waals gap, the (00$l$) diffraction peaks in Tl$_{0.6}$Bi$_2$Te$_3$ shift to higher 
2$\theta$ angles, as one can clearly see in the inset of Figure 1(a). This means that the
lattice parameter along the $c$-axis gets {\it shorter} after Tl
doping. Quantitatively, it decreases from 30.606(4) {\AA} in
Bi$_2$Te$_3$ to 30.438(9) {\AA} in Tl$_{0.6}$Bi$_2$Te$_3$. 
This observation suggests that intercalation is not taking place in 
Tl$_{0.6}$Bi$_2$Te$_3$. Note that the ICP-AES analysis indicates the 
existence of nearly stoichiometric amount of Tl in superconducting crystals, 
as shown in Table S1 of Supporting Information.

We also measured powder XRD patterns of Tl$_{0.6}$Bi$_2$Te$_3$ 
with Cu $K_{\alpha}$ radiation in Bragg-Brentano geometry on powders 
obtained from crushed crystals, and the results are 
shown in Figure 1(b) along with a Rietveld refinement. 
(Similar XRD data for smaller Tl contents are shown  
in Figure S1 of the Supporting Information without refinements.)
We note, however, that after the grinding, the powdered samples are no longer superconducting.
This suggests that the superconductor phase of Tl$_{0.6}$Bi$_2$Te$_3$ is unstable and is fragile against mechanical 
strain.
Furthermore, we observed that the superconducting volume fraction in Tl$_{0.6}$Bi$_2$Te$_3$
diminishes with time when the samples are left at room temperature, 
even though they are kept in inert atmosphere or in vacuum; this suggests that doped Tl atoms are 
mobile even at room temperature.
In passing, we have also tried to perform single-crystal XRD analysis, but Tl$_{0.6}$Bi$_2$Te$_3$ is so soft 
that preparations of small single crystals required for this kind of analysis resulted in deformed samples, making
it impossible to obtain data of sufficient quality for the crystal structure analysis. 
The degradation of crystal quality was also apparent in powdered samples.

As one can see in Figure 1(b), the diffraction data are
well described by two coexisting phases, Bi$_2$Te$_3$ and TlBiTe$_2$, when
taking the preferred orientation correction into account.  The
TlBiTe$_2$ phase possesses a volume fraction of about 35\%. Attempts to
refine the occupation of the Tl ions at the intercalation or other interstitial positions
in the Bi$_2$Te$_3$ phase did not yield a significant occupation,
in agreement with the observation that the $c$ lattice parameter
is shorter than that in pristine Bi$_2$Te$_3$. We find a significant
amount of vacancies on the Bi site (about one third), which
indicates a massive occupation of the Te sites by Bi or Tl ions (i.e. 
Bi$^{'}_{\rm Te}$ or Tl$^{'''}_{\rm Te}$ antisite defects).
Note that Bi and Tl are indistinguishable in X-ray diffraction due to their similar atomic
numbers.
For the Rietveld refinement, the structure of Bi$_2$Te$_3$ with
symmetry $\mathrm{R}\bar{3}\mathrm{m}$ (lattice constants
$a$ = $b$ = 4.3850(16) \AA , $c$ = 30.438(9) \AA, and $\gamma$ = 120$^\circ$)
with additional Tl positions was used. The position of the
Bi-atoms was refined to $(0, 0, 0.3988(3))$ at the Wyckoff
position 6c, the positions of the Te-atoms were $(0, 0, 0)$ at the
Wyckoff position 3a and $(0, 0, 0.8043(2))$ at the Wyckoff
position 6c. No significant occupation of additional Tl-atoms at
$(0.5, 0, 0.5)$ or $(0.0, 0, 0.5)$ could be determined. All
positions refer to the hexagonal setting of the rhombohedric cell.
The TlBiTe$_2$ impurity phase was also described in space group
$\mathrm{R}\bar{3}\mathrm{m}$ (lattice constants $a$ = $b$ = 4.539(1)
\AA , $c$ = 22.617(8) \AA, and $\gamma$ = 120$^\circ$) and only the
$z$ position of the Te site was refined to $(0, 0, 0.2446(10))$.


Although TlBiTe$_2$ was reported to become superconducting below
0.14 K, \cite{TlBiTe2_SC} this impurity phase cannot be responsible for
the appearance of the superconductivity in our samples, whose $T_{c}$ is
above 2 K. 
It is also worth mentioning that elemental thallium metal is superconducting with
$T_c$ of $\sim$2.4 K,\cite{Tl-SC} which is close to the $T_c$ of Tl$_{0.6}$Bi$_2$Te$_3$. However, it is very unlikely that the superconductivity observed here is due to elemental thallium, because the XRD data do not indicate the existence of thallium metal in our samples.

In the past, the crystal structure of Tl-doped Bi$_2$Te$_3$ with the composition of Tl$_x$Bi$_{2-x}$Te$_3$ was studied.\cite{Tl-BT1,Tl-BT2} It was concluded that, even though the composition suggests that Tl atoms partially substitute the Bi sites of the Bi$_2$Te$_3$ lattice, what actually happens is that Tl nucleates microscopic patches of nominal Te-Bi-Te-Tl-$V^{\bullet\bullet}_{\rm Te}$ layer, which is derived from TlBiTe$_2$ structure and has the same symmetry as the Bi$_2$Te$_3$ phase \cite{Tl-BT1,Tl-BT2} (in real crystals, the fictitious plane of Te vacancy would be partially filled with Te, distributing $V^{\bullet\bullet}_{\rm Te}$ to the neighboring Te plane of Bi$_2$Te$_3$). It was proposed that there are random microscopic formations of this defect layer in Tl-doped Bi$_2$Te$_3$, which results in the overall crystal structure to be the same as Bi$_2$Te$_3$ and causes little change in the lattice constants, even though a significant amount of Tl is incorporated into the lattice. 


It is useful to note that both our ICP-AES and EDX analyses of the crystals indicate the presence of nearly stoichiometric amount of Tl, which would give rise to about 30\% of the TlBiTe$_2$ phase if the sample phase-separates into Bi$_2$Te$_3$ and TlBiTe$_2$. The amount of the TlBiTe$_2$ phase indicated in the Rietveld refinement is consistent with this estimate, which suggests that due to the mobility of Tl atoms at room temperature, the material actually phase separates into Bi$_2$Te$_3$ and TlBiTe$_2$ upon grinding. This in turn suggests that it is very difficult to elucidate the crystal structure of the superconductor phase.

A possible picture, which one can speculate for the superconducting phase based on the above result, would be to consider the formation of the nominal Te-Bi-Te-Tl-$V^{\bullet\bullet}_{\rm Te}$ defect layer in the Bi$_2$Te$_3$ lattice, as is the case of the Tl$_x$Bi$_{2-x}$Te$_3$ compound.\cite{Tl-BT1,Tl-BT2}. This defect layer may eventually cluster to form the TlBiTe$_2$ phase. An important difference from the case of the Tl$_x$Bi$_{2-x}$Te$_3$ compound would be that a sizable portion of Bi atoms in Tl$_{0.6}$Bi$_2$Te$_3$ are most likely partially filling the Te sites of the Bi$_2$Te$_3$ lattice and form Bi$^{'}_{\rm Te}$ antisite defects, which is consistent with the result of the Rietveld refinement. In fact, the composition of Tl$_{0.6}$Bi$_2$Te$_3$ would create a significantly Te-deficient growth condition and promote the formation of Bi$^{'}_{\rm Te}$ antisite defects.\cite{Scanlon} In any case, the precise structure of superconducting Tl$_{0.6}$Bi$_2$Te$_3$ should be determined in future studies, possibly by neutron scattering on as-grown crystals.

\begin{figure}
\includegraphics[width=7.5cm,clip]{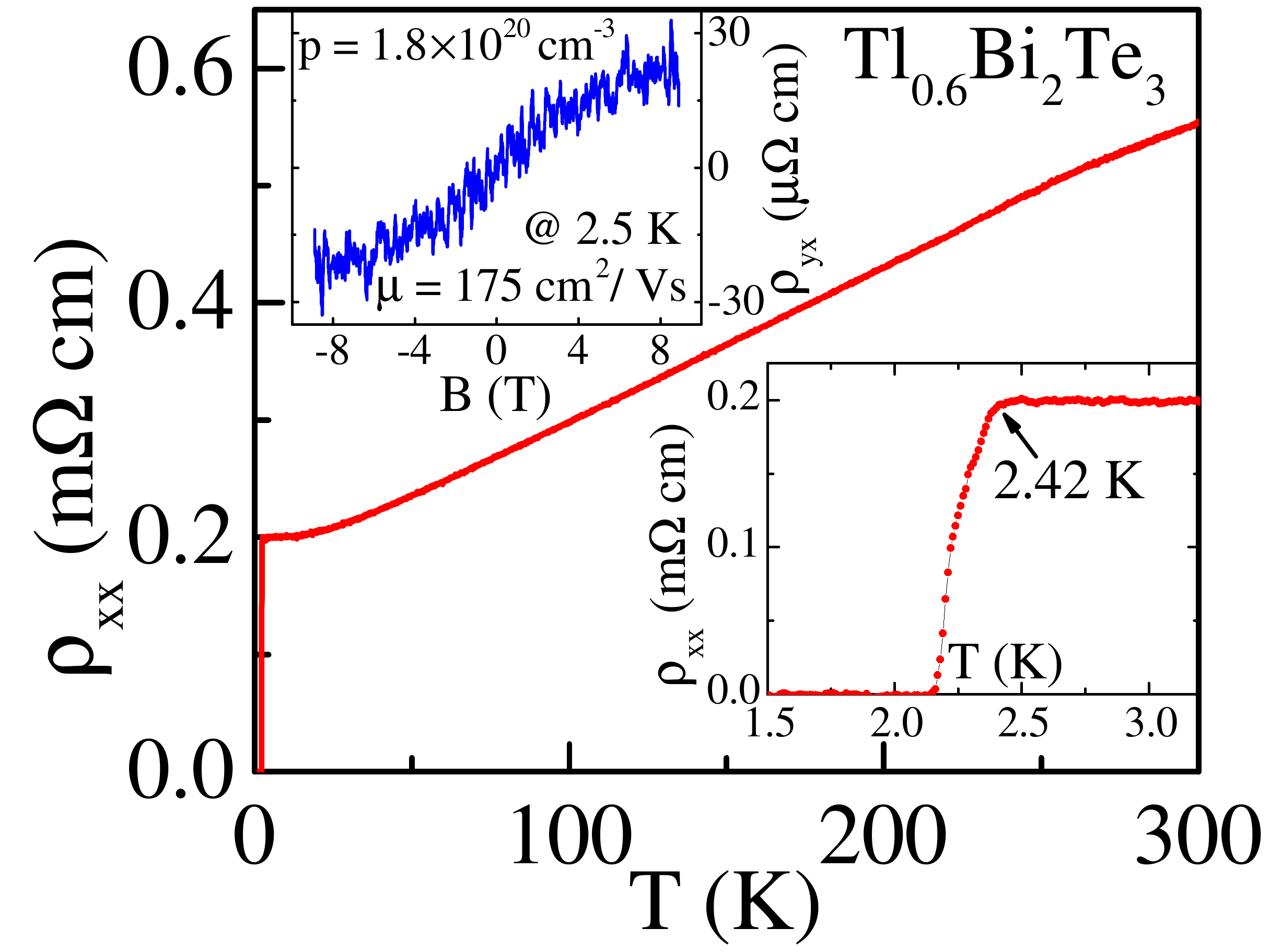}
\caption{
Temperature dependence of the in-plane resistivity $\rho_{xx}$ in
Tl$_{0.6}$Bi$_2$Te$_3$. Upper inset shows the magnetic-field dependence
of the Hall resistivity $\rho_{yx}$ of the same sample at 2.5 K; lower
inset shows the $\rho_{xx}$($T$) behavior near the transition. 
}
\label{Transport_plot}
\end{figure}

Figure 2 shows the temperature dependence of $\rho_{xx}$ in
Tl$_{0.6}$Bi$_2$Te$_3$ at zero field. The onset of superconducting
transition occurs at $T \approx 2.42$ K, and the zero resistivity is
achieved at $T \approx 2.15$ K (lower inset of Figure 2), indicating a
relatively sharp transition. The resistivity in the normal state shows a
metallic behavior with the residual resistivity $\rho_0$ = 2 $\times$
10$^{-4}$ $\Omega$cm. The magnetic-field dependence of $\rho_{yx}$
at 2.5 K is shown in the upper inset of Figure 2; this
$\rho_{yx}(B)$ behavior is slightly non-linear, which suggests the
existence of two or more bands at the Fermi level. Also, the
$\rho_{yx}(B)$ data indicate that the main carriers are $p$-type (i.e.
holes), and from the slope near 0 T we calculate the approximate carrier
density of $p \approx {1.8} \times 10^{20}\,{\rm cm}^{-3}$. From $p$ and
$\rho_0$, one obtains the mobility $\mu \approx$ 175 cm$^{2}$/Vs. 
It is important to note that the carrier type is different from the case of
Cu- or Sr-intercalated Bi$_2$Se$_3$ superconductors, in which the
carriers are $n$-type. \cite{Hor_PRL_10, Markus_PRL_11, SrxBi2Se3_15}
Nevertheless, the magnitude of the carrier density, about 2 $\times$
10$^{20}$ cm$^{-3}$, is comparable to that in Cu$_x$Bi$_2$Se$_3$.
\cite{Hor_PRL_10, Markus_PRL_11} Hence, Tl$_{0.6}$Bi$_2$Te$_3$ would
allow for investigation of the roles of the carrier types in producing a
topological superconducting state in otherwise similar settings, if this
material turns out to be a TSC. In passing, we comment on the possible impact of 
the TlBiTe$_2$ impurity phase and the nominal Te-Bi-Te-Tl-$V^{\bullet\bullet}_{\rm Te}$ 
defect layer on the transport properties. While the direct impact of 
phase-separated TlBiTe$_2$ impurity phase is expected to be
minor because the carrier density of this phase is similar to that of the main phase,\cite{TlBiTe2_SC}
the defect layer may be working as strong scatters of charge carriers and is possibly playing 
some role in the occurrence of superconductivity.

\begin{figure*}
\includegraphics[width=12cm,clip]{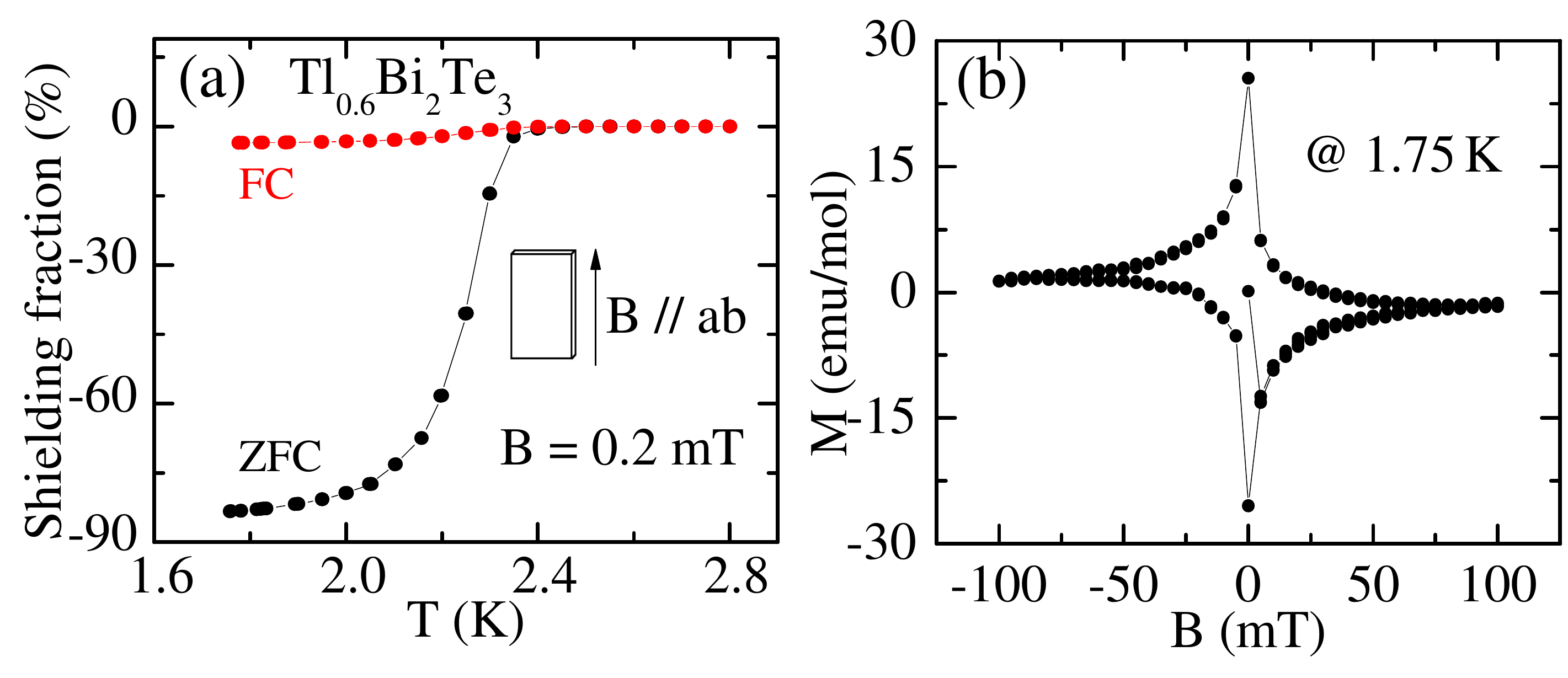}
\caption{
(a) Temperature dependence of the magnetic susceptibility in
Tl$_{0.6}$Bi$_2$Te$_3$ under 0.2 mT for FC and ZFC measurements, 
plotted in terms of the shielding fraction. 
(b) Plot of magnetization $M$ vs magnetic field $B$ at 1.75 K. 
}
\label{magnetization_plot}
\end{figure*}

Figure 3(a) shows the temperature dependence of the shielding fraction
in Tl$_{0.6}$Bi$_2$Te$_3$ measured under 0.2 mT applied
parallel to the $ab$-plane to minimize the demagnetization effect; the
configuration is schematically shown in the inset. Note that the
shielding fraction is defined as the fraction of the sample volume from
which the magnetic field is kept out due to superconductivity; the data for
both field-cooled (FC) and zero-field-cooled (ZFC) measurements are shown. 
The onset of superconducting transition is observed
at $T \simeq$ 2.35 K. This is consistent with the resistivity transition shown 
in Figure 2. Furthermore, the ZFC shielding fraction at 1.75 K is 
as much as 83\%, pointing to bulk superconductivity. 

We have also synthesized
Tl$_x$Bi$_2$Te$_3$ samples with various $x$ values, and
it was found that both $T_c$ and the shielding fraction become lower for 
$x <$ 0.6, as is shown in Figure S2 of the Supporting Information. 
Also, for $x >$ 0.6, we found that the TlBiTe$_2$ impurity phase
becomes dominant and it was impossible to synthesize large single
crystals retaining the Bi$_2$Te$_3$ structure. Therefore, we concluded
that $x$ = 0.6 is the optimum composition for this new superconductor.

The magnetization curve $M(B)$ measured at 1.75 K with the
magnetic field applied parallel to the $ab$ plane is shown in Figure
3(b). This $M(B)$ behavior indicates that Tl$_{0.6}$Bi$_2$Te$_3$ is a type-II
superconductor and the flux pinning is very weak, as was also the case
in Cu$_x$Bi$_2$Se$_3$. \cite{Markus_PRL_11} From the low-field $M(B)$
behavior measured after zero-field cooling (shown in Figure S3 of
Supporting Information), one can determine the lower critical field
$B_{c1}$ as the characteristic field above which the $M(B)$ data start
to deviates from the initial linear behavior; at the lowest temperature
of 1.75 K, $B_{c1}$ is estimated to be 0.35 mT, which is very small and
is comparable to that in Cu$_x$Bi$_2$Se$_3$. \cite{Hor_PRL_10,
Markus_PRL_11} Such a low $B_{c1}$ value means a very low superfluid
density, which is consistent with the low carrier density.

\begin{figure}[b]
\includegraphics[width=8.9cm,clip]{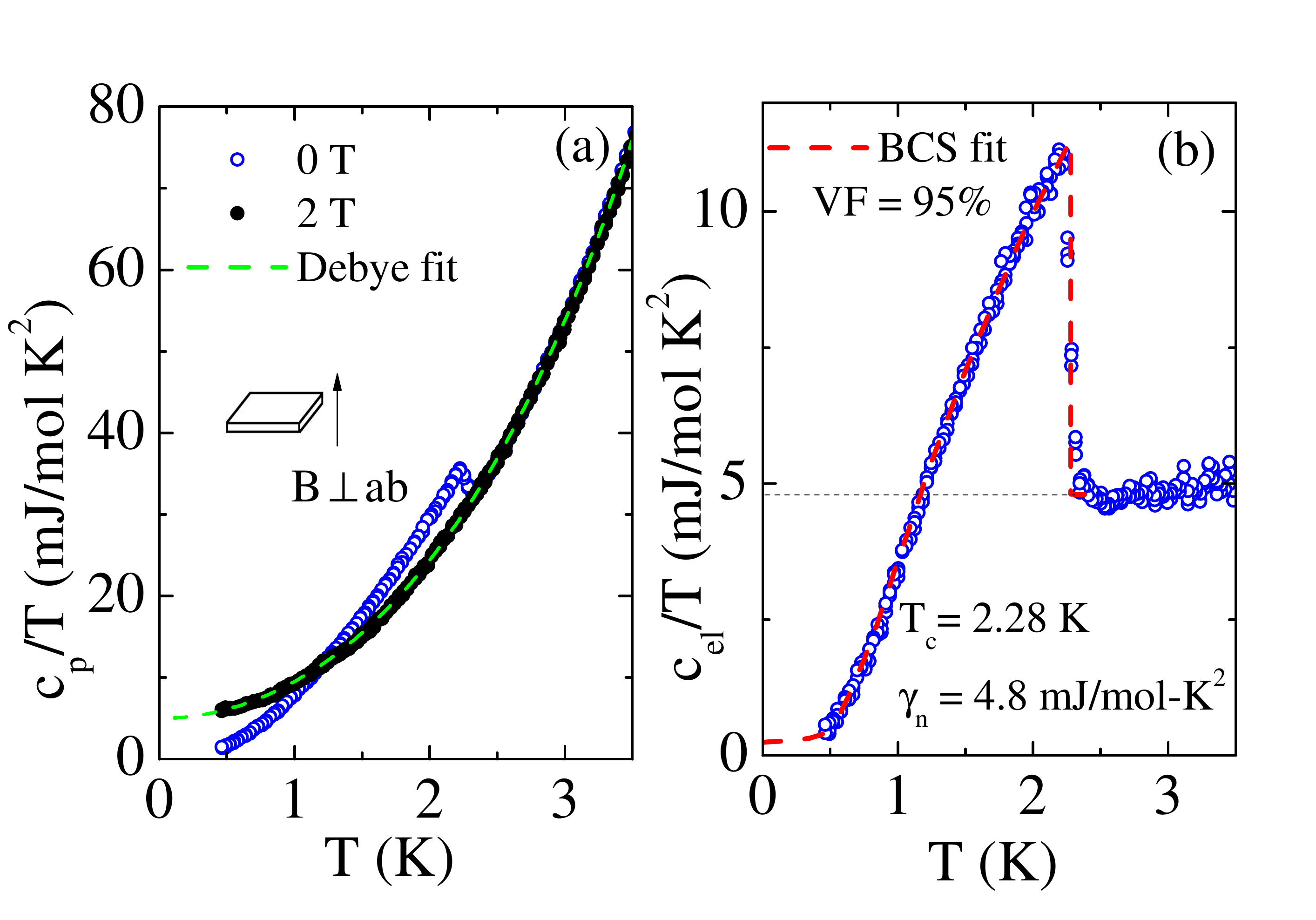}
\caption{
(a) Plots of $c_p (T) / T$ vs $T$ for Tl$_{0.6}$Bi$_2$Te$_3$
measured in 0 and 2 T applied along the $c$ axis; dashed line
shows the conventional Debye fitting. (b)
The electronic contribution $c_{el} / T$ in 0 T obtained 
after subtracting the phonon term determined in 2 T;
dashed line shows a BCS-model fitting assuming 95\%
superconducting volume fraction. 
}
\label{specific heat_plot}
\end{figure}

Figure 4 shows the plots of $c_{p}/T$ vs $T$ measured in 0 T and 2 T
applied perpendicular to the $ab$-plane, as schematically shown in the inset; since the superconductivity is
completely suppressed in 2 T as we show later, the 2-T data represent
the normal-state behavior. A fitting of the normal-state data to the
conventional Debye formula $c_{p} = \gamma_{n}T + A_{3}T^{3} +
A_{5}T^{5}$, shown as the dashed line in Figure 4(a), gives the
following parameters: $\gamma_{n}$ = 4.8 mJ/mol-K$^{2}$, $A_{3}$ = 4.4
mJ/mol-K$^{4}$, and $A_{5}$ = 0.11 mJ/mol-K$^{6}$. The electronic
specific heat $c_{el}/T$ in the SC state is obtained by subtracting the
phononic contribution $A_{3}T^{3} + A_{5}T^{5}$ from the zero-field
data, and the result is plotted in Figure 4(b). The pronounced jump gives 
evidence for the bulk nature of the superconductivity in
Tl$_{0.6}$Bi$_{2}$Te$_{3}$, and this anomaly provides an accurate measure of
$T_c$ = 2.28 K. Fitting of $c_{el}(T)/T$ to the BCS model
\cite{BCS} reproduces the zero-field data very well if one assumes a
95\% superconducting VF. Therefore, one may conclude that the
superconducting state of Tl$_{0.6}$Bi$_{2}$Te$_{3}$ is fully gapped. Note
that the applicability of the BCS model to the specific-heat data does
not exclude the possibility of unconventional odd-parity pairing.
\cite{Markus_PRL_11} The superconducting VF of 95\% is incompatible with 
the 35\% inclusion of TlBiTe$_2$ phase suggested by the Rietveld analysis 
on crushed crystals, and this incompatibility supports our speculation that 
a sizable amount of TlBiTe$_2$ phase is created upon grinding.

\begin{figure*}[t]
\includegraphics[width=18cm,clip]{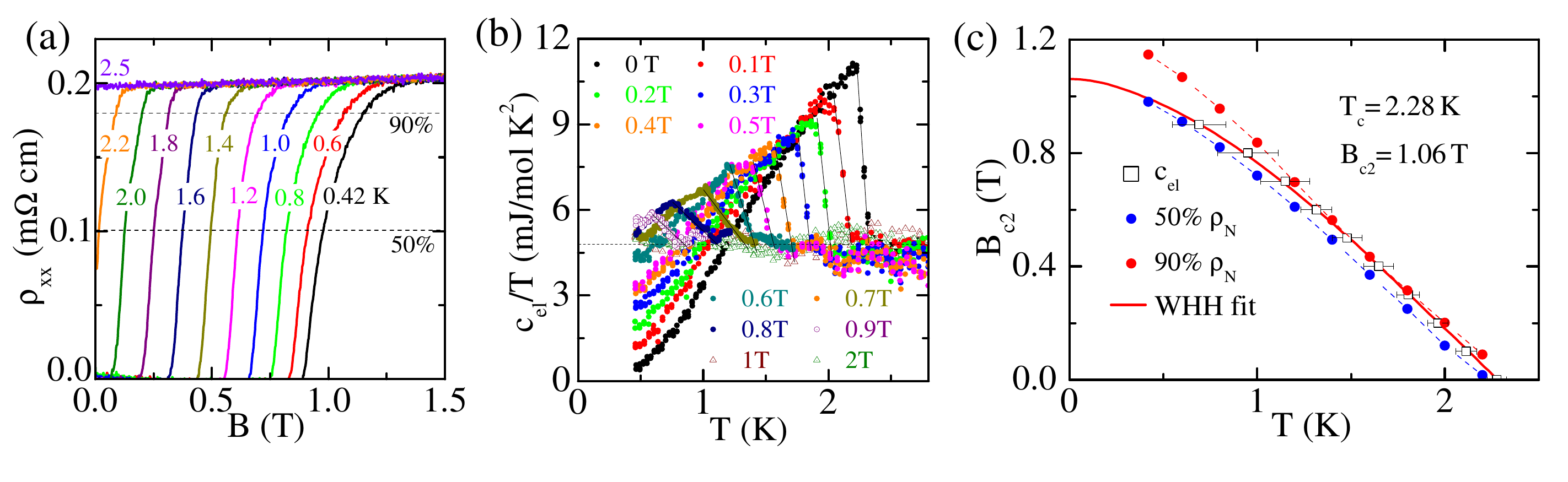}
\caption{
(a) $\rho_{xx}(B)$ curves for Tl$_{0.6}$Bi$_2$Te$_3$ in the transition region 
at various temperatures. The magnetic-field direction
is perpendicular to the $ab$ plane. 
(b) Plots of $c_{el} (T) / T$ vs $T$ in various magnitudes of perpendicular
magnetic field. (c) $B_{c2}$ vs $T$ phase diagram determined from 
50\% $\rho_N$ (red circles), 90\% $\rho_N$ (blue circles), and $c_{el}$ (black squares) data;
the error bar on the data points from $c_{el}$ corresponds to the width of the specific-heat jump. 
The solid line shows the WHH fitting to the thermodynamic $B_{c2}(T)$ obtained
from $c_{el}$.
}
\label{Hc2_plot}
\end{figure*}

To determine the upper critical field $B_{c2}$, the magnetic-field
dependences of $\rho_{xx}$ at various
temperatures down to 0.42 K were measured in fields perpendicular to
the $ab$-plane [Figure 5(a)]. For the analysis of the resistive
transitions, both the 50\% and 90\% levels of the normal-state resistivity $\rho_N$
(shown by dashed lines) are taken as characteristic levels to mark
the transition; the difference between these two criteria gives an idea about the uncertainly 
in determining $B_{c2}$ from resistive transitions.
In addition, the $c_{el}(T)/T$
behavior was measured in various magnetic-field strengths [Figure 5(b)],
and we take the mid-point of the specific-heat jump as the definition of
the thermodynamic transition. 
Note that the data shown in Figures 2 -- 5 are all 
taken on the same sample. The summary of $B_{c2}$ thus determined
are plotted in Figure 5(c). The Werthamer-Helfand-Hohenberg (WHH) theory
\cite {WHH} fits the thermodynamic $B_{c2}(T)$ obtained from specific
heat very well and gives $B_{c2}(0)$ of 1.06 T, which corresponds to the
coherence length $\xi = \sqrt{\Phi_0/(2\pi B_{c2})}$ = 17.6 nm. On the
other hand, the $B_{c2}(T)$ extracted from resistive transitions do not
follow the WHH behavior and extrapolates to a higher $B_{c2}(0)$; such a
behavior has been reported for Cu$_x$Bi$_2$Se$_3$ and also for
pressurized Bi$_2$Se$_3$, and was argued as evidence for unconventional
superconductivity.\cite{HP-Bi2Se3_PRL, CuBi2Se3Hc2}

\section{Conclusions}

The discovery of superconductivity in Tl$_{0.6}$Bi$_2$Te$_3$ widens the
opportunities to elucidate topological superconductivity in
topological-insulator-based superconductors, particularly since the
superconducting VF of up to 95\% is achievable. Various aspects of the
superconductivity in Tl$_{0.6}$Bi$_2$Te$_3$, including the unconventional
resistive $B_{c2}(T)$ behavior and the very small $B_{c1}$ value, are
similar to those found in Cu$_x$Bi$_2$Se$_3$. Nevertheless, the carrier
type is opposite, which may prove useful for understanding the mechanism
of superconductivity. The crystal structure of this material appears to be 
essentially unchanged from that of Bi$_2$Te$_3$ with a slightly shorter
$c$-axis length and no interstitials, but it turned out to be difficult to 
elucidate the exact structure of the superconductor phase.

\section{Supporting Information}

Table showing the results of ICP-AES analysis; 
powder XRD data for smaller Tl contents;
superconducting transitions in crystals with smaller Tl contents 
probed by magnetic susceptibility; 
virgin $M(B)$ curve for determining $B_{c1}$.

\section{Acknowledgment}

This work was supported by Japan Society for the Promotion of Science
(KAKENHI 25220708 and 25400328) and the Excellence Initiative of the
German Research Foundation.



\clearpage

\renewcommand{\thefigure}{S\arabic{figure}} 

\setcounter{figure}{0}

\renewcommand{\thesection}{S\arabic{section}.}

\begin{flushleft} 
{\Large {\bf Supporting Information}}
\end{flushleft} 

\vspace{2mm}



\begin{table}[b]
\centering 
\begin{tabular}{c c c c c } 
\hline\hline 
$x$ & Nominal composition & Tl &  Bi &  Te   \\
[0.5ex] 
\hline 
0.1	& Tl$_{0.1}$Bi$_{2}$Te$_{3}$ & 0.106(1) & 2.087(6) & 3 \\
0.2	& Tl$_{0.2}$Bi$_{2}$Te$_{3}$ & 0.207(2) & 2.098(6) & 3 \\
0.3	& Tl$_{0.3}$Bi$_{2}$Te$_{3}$ & 0.233(2) & 2.090(6) & 3 \\
0.4	& Tl$_{0.4}$Bi$_{2}$Te$_{3}$ & 0.396(4) & 2.077(6) & 3 \\
0.5 & Tl$_{0.5}$Bi$_{2}$Te$_{3}$ & 0.526(5) & 2.040(6) & 3  \\
0.6 & Tl$_{0.6}$Bi$_{2}$Te$_{3}$  & 0.609(6) & 2.059(6) & 3  \\
 [1ex] 
\hline 
\end{tabular}
\vspace{5mm}
\caption{{\bf Molar ratio of Tl, Bi, and Te in the single-crystal
samples of Tl$_x$Bi$_2$Te$_3$.} 
The data are obtained from ICP-AES analyses. The tellurium composition is fixed to be 3.
} 
\end{table}


\begin{figure}[b]
\includegraphics[width=11cm,clip]{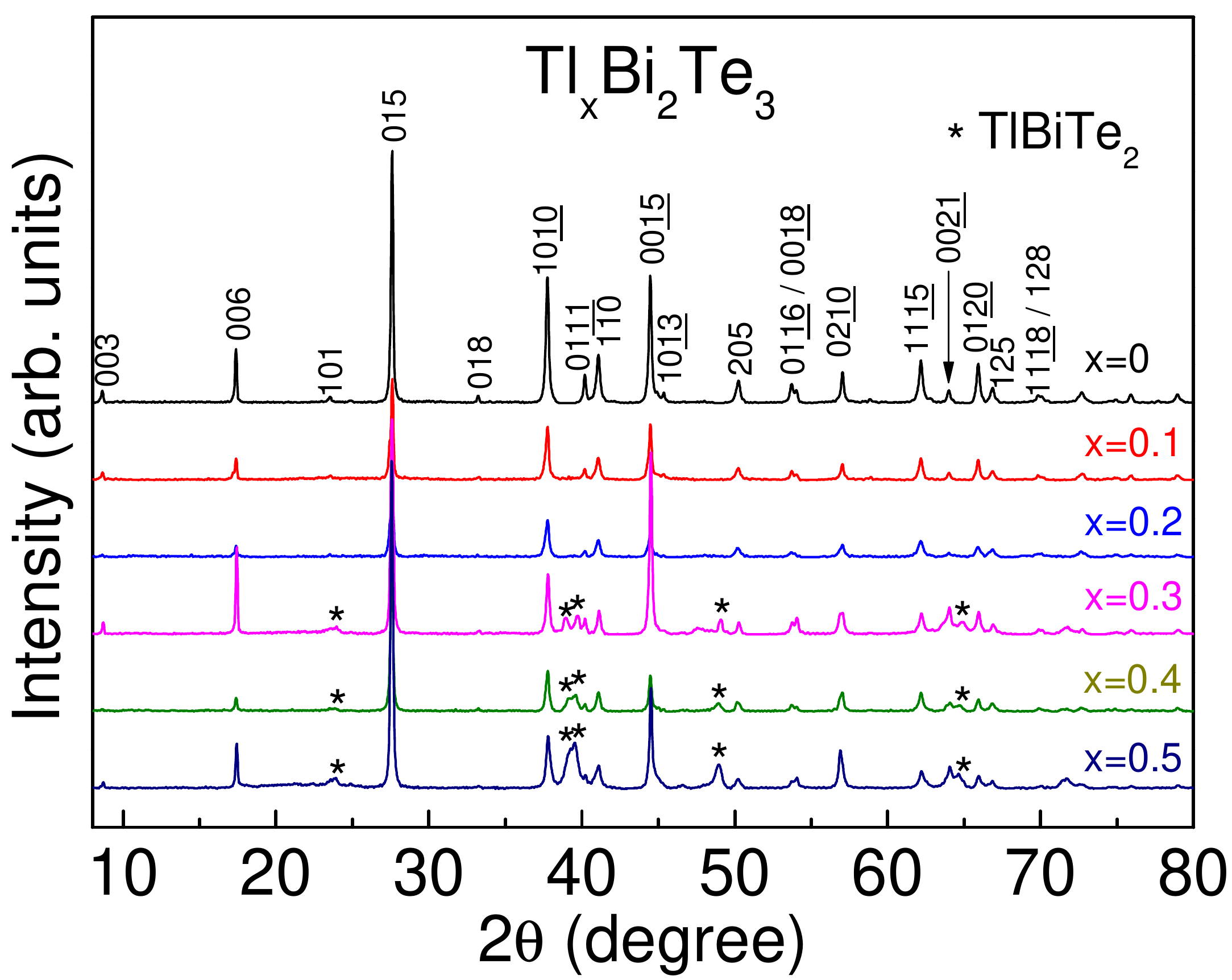}
\caption{ Powder XRD patterns of Tl$_{x}$Bi$_2$Te$_3$ ($x$ = 0.1, 0.2, 0.3,
0.4, and 0.5) and Bi$_2$Te$_3$, 
taken on powders prepared by crushing cleaved single crystals. 
The peaks due to the TlBiTe$_2$ impurity phase are indicated with asterisks, 
and they are discernible in the data for $x$ = 0.3, 0.4, and 0.5.} 
\end{figure}

\begin{figure}
\includegraphics[width=8.5cm,clip]{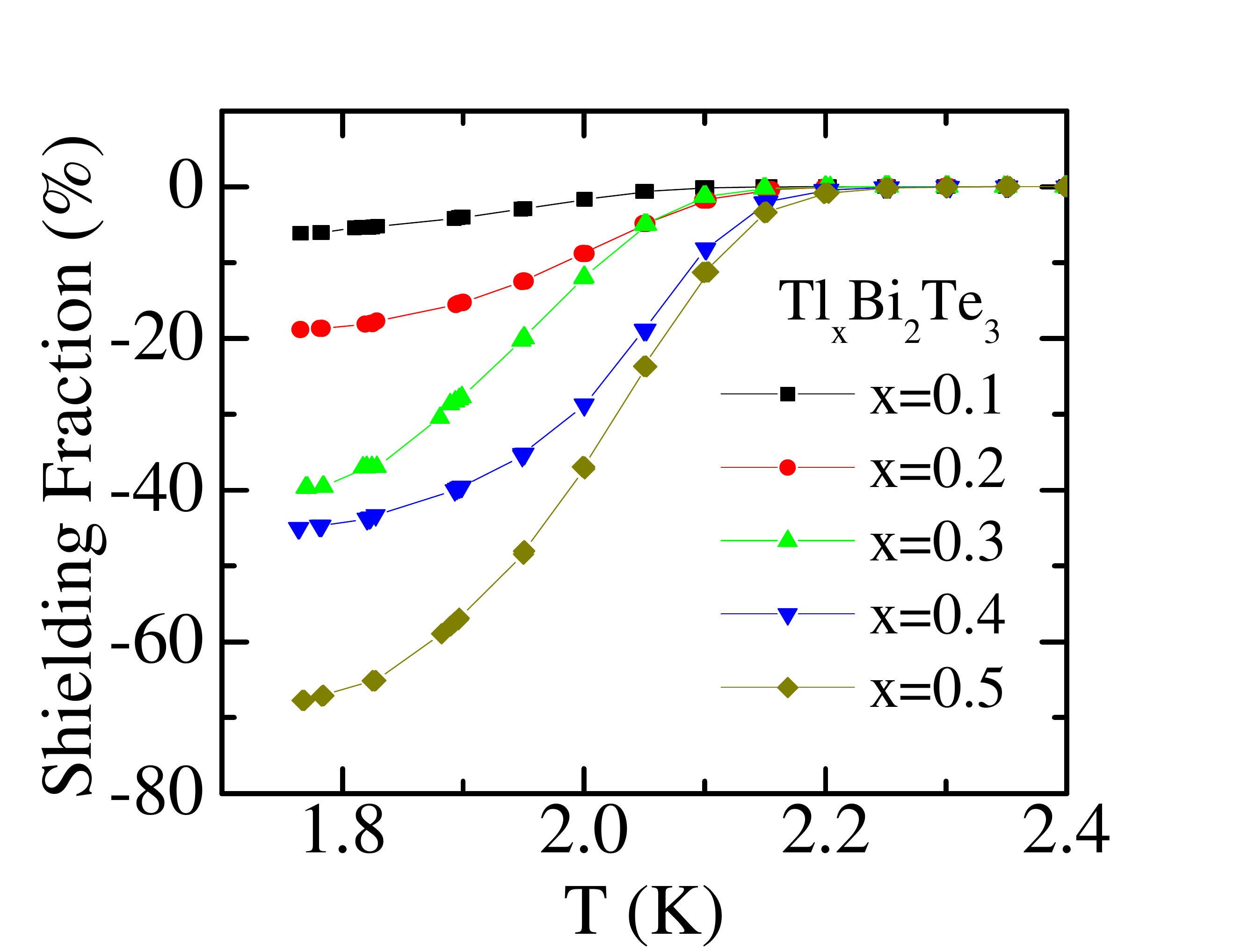}
\caption{
Temperature dependence of the magnetic susceptibility in
Tl$_x$Bi$_2$Te$_3$ with $x$ = 0.1 -- 0.5 measured under 0.2 mT, 
plotted in terms of the shielding fraction. 
} 
\end{figure}

\begin{figure}
\includegraphics[width=6.5cm,clip]{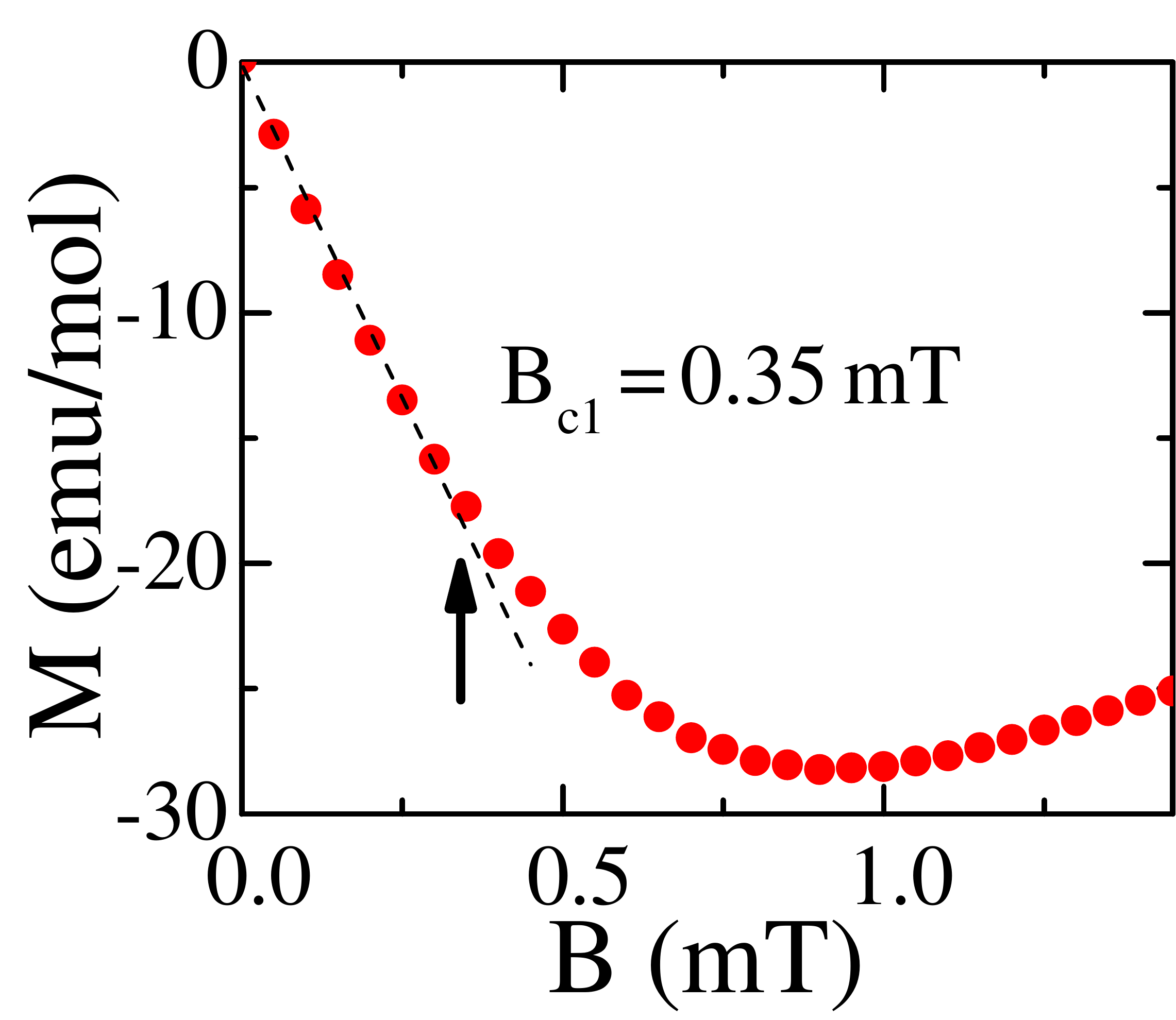}
\caption{Virgin $M(B)$ curve of Tl$_{0.6}$Bi$_2$Te$_3$ at 1.75 K measured after 
zero-field cooling the sample. The magnetic field was applied parallel to the 
$ab$ plane to minimize the demagnetization effect.
} 
\end{figure}

\end{document}